\documentclass[twocolumn]{openjournal}
\usepackage{graphicx}   
\usepackage{amsmath}    
\usepackage{amssymb,amsfonts}    

\usepackage[T1]{fontenc}
\usepackage{ae,aecompl}

\usepackage{natbib}   

\shorttitle{Pulsation in hot main sequence stars}
\shortauthors{L. A. Balona}

\begin{document}

\title{Pulsation in hot main sequence stars: comparison of observations with
models}

\email{email: lab@saao.ac.za}

\author{L. A. Balona}

\affiliation{South African Astronomical Observatory, P.O. Box 9, Observatory
7935, Cape Town, South Africa}

\begin{abstract}
Comparison of {\em TESS} observations with models of pulsating hot main 
sequence stars reveals several problems.  The frequencies in $\delta$~Scuti 
stars at a given temperature do not match those predicted by models.  A 
large fraction of $\delta$~Scuti stars are hotter than the predicted hot edge 
of the instability strip. These Maia variables occupy a region in the H--R
diagram between the $\delta$~Scuti and $\beta$~Cephei stars.  Their high 
frequencies cannot be explained by rapid rotation.  The $\gamma$~Doradus stars 
are to be found together with $\delta$~Scuti stars all along the main sequence 
where they merge with SPB stars.  They are an example of the extraordinary 
frequency pattern variations seen in $\delta$~Scuti stars, suggesting that the 
linear behaviour assumed in the models is incorrect.  A well-defined upper 
envelope in frequency as a function of effective temperature is found in 
$\delta$~Scuti and Maia variables.  The frequencies of $\beta$~Cephei stars are
generally higher than predicted. The gap in the \mbox{H--R} diagram between 
$\beta$~Cephei and hot SPB stars predicted by the models is not observed.  Most 
$\beta$~Cephei variables contain low frequencies typical of SPB stars. There are 
no discernible boundaries between the traditional classes of pulsating stars. 
A major revision of pulsation in upper main sequence stars is required.
\end{abstract}


\section{Introduction}

Prior to space observations, stellar pulsation among hot main sequence stars
appeared to be well understood.  The $\delta$~Scuti (DSCT) variables are 
F5--A2 dwarfs or giants with frequencies 5--50\,d$^{-1}$ in which pulsational 
driving is mostly attributed to the opacity $\kappa$~mechanism operating in the
He\,{\sc II} partial ionization zone.  For the cooler DSCT stars, coupling 
between convection and pulsation is important and affords an explanation for 
the $\gamma$~Doradus (GDOR) variables \citep{Dupret2005, Xiong2016}.  These are 
stars with pulsation frequencies less than 5\,d$^{-1}$ lying on the cool edge 
of the DSCT instability strip.

The $\beta$~Cephei (BCEP) variables, which are B4--O9 main sequence stars 
pulsating with frequencies in the range 3--20\,d$^{-1}$, are driven by the 
opacity mechanism operating in the partial ionization zone of iron-like metals.
The same mechanism is responsible for the SPB stars which have frequencies less
than 3\,d$^{-1}$.  In these less massive stars, the iron opacity bump is 
located in a deeper layer. Together with the lower luminosity, this leads to an
increase in the thermal timescale and lower pulsation frequencies.  These
SPB stars partially overlap the BCEP variables and extend to spectral type B8.  
Between the cool edge of the SPB variables and the hot edge of the DSCT stars, 
models do not predict pulsation.

Photometric time-series observations from space missions such as {\em CoRoT,
Kepler} and {\em TESS} have increasingly challenged the above perceptions.
From {\em CoRoT} observations, \citet{Degroote2009b} reported low-amplitude 
late B-type pulsators lying between the SPB and DSCT instability strips with 
frequencies similar to those in DSCT stars. Such ``Maia'' variables, as they 
have historically been called, had been suspected from ground-based photometry 
\citep{McNamara1985,Lehmann1995,Percy2000,Kallinger2004}. The existence of 
MAIA variables is now well established, not only from {\em Kepler} and 
{\em TESS} missions \citep{Balona2020a, Balona2023a}, but also from {\em Gaia} 
photometry \citep{Gaia2023}. 

The general view is that MAIA variables do not constitute a new group of
pulsating stars.  The high frequencies are presumed to be gravito-inertial 
modes shifted to high frequencies in rapidly rotating SPB stars 
\citep{Townsend2005d, Salmon2014}.  However, an analysis of their projected 
rotational velocities shows that MAIA stars have the same rotational
velocity distribution as main sequence stars.  In addition, over 10\,\% 
of MAIA variables have frequencies exceeding 60\,d$^{-1}$, which certainly
cannot be a result of rotation.  The MAIA variables appear to be an extension 
of the DSCT variables to early B stars \citep{Balona2023a}.

Gravito-inertial modes have been detected in several stars 
\citep{Papics2012, VanReeth2018, Mombarg2019, Ouazzani2020}. These seem 
to be present, together with acoustic gravity modes, in GDOR stars. They
arise in rotating stars where the Coriolis force acts as a restoring force.
Rossby waves are seen as broad features in the periodograms of some DSCT 
variables \citep{Saio2018a}. \citet{Mirouh2022} has presented an excellent
review of the various types of oscillations that may be present in rotating
stars.

Models do not predict low frequencies for main sequence stars with effective 
temperatures in the range $7500 < T_{\rm eff} < 11\,000$\,K.  Yet there are a
substantial number of GDOR stars hotter than $7500$\,K \citep{Balona2016c}.   
The GDOR and SPB stars form a continuous group of low-frequency
variables.  Whether or not pulsation in these hot GDOR stars may be attributed 
to inertial modes is an open question.  Low frequencies are also present in 
most DSCT stars hotter than about 7500\,K \citep{Grigahcene2010}.  These used 
to be known as DSCT+GDOR ``hybrids'', but since most DSCT stars are hybrids,
the term is no longer useful.

The rapidly-oscillating Ap (roAp) stars are Ap/Fp main sequence stars that
pulsate in a limited high-frequency range above 60\,d$^{-1}$
\citep{Kurtz1982}.  Until recently, such high frequencies were assumed to be
confined to chemically peculiar stars, but data from the {\em TESS} mission has
shown that similar frequencies occur in many chemically normal stars
\citep{Balona2022a}.  It appears that roAp stars should not be regarded as
a separate class but as ordinary members of the DSCT class.

The mass loss in Be stars is frequently attributed to pulsation.  However, the 
quasi-periodic light variations are not coherent, as expected from a pulsating 
star. There is a characteristic pattern in the light curves and periodograms of
Be stars which can be understood as a result of co-rotating circumstellar 
clouds \citep{Balona2020b, Balona2021d}.  Be stars are not considered in this 
paper. 

It turns out that there are no clear boundaries between any of the several 
types of hot pulsating stars GDOR, DSCT, MAIA, SPB and BCEP
\citep{Balona2020a}.  This complicates the definition of the variability 
classes because arbitrary boundaries in effective temperature and frequency 
need to be introduced to uniquely define each variability class.

A very strange aspect is the huge variety of frequency patterns to be found in 
DSCT stars with similar effective temperatures, luminosities, and rotation rates 
\citep{Balona2024a}.  Stars with similar parameters should pulsate with 
similar frequencies and amplitudes, as models predict.  This appears not to be 
the case.  Each DSCT star has a unique frequency pattern.  

Non-adiabatic pulsation models assume that pulsation is a linear
phenomenon at very low amplitudes.  A mode that is unstable at a given
frequency is assumed to grow to a limiting amplitude while maintaining
nearly the same pulsation frequency.  It seems that this fundamental 
assumption is not correct for DSCT stars (and perhaps for other classes 
as well).  As a result, the frequency that is observed is probably a 
result of a non-linear interaction and bears no resemblance to the 
frequency predicted by the models \citep{Mourabit2023}.

It seems that current perceptions of stellar pulsation in hot main
sequence stars, which were adequate to explain ground-based observations, are 
no longer sufficient.  This paper aims to compare the predicted
location of pulsation instability and the frequencies with {\em TESS}
observations of hot main sequence stars. In this way, one might obtain clues as
to how the models could be modified to better agree with observations. This 
work is based on a catalogue of the variability classes of over 125\,000 
{\em Kepler} and {\em TESS} stars on the upper main sequence compiled by 
\citet{Balona2022c} which is freely available.

\section{The data and variability classification}

The light curves obtained from the {\em Kepler} and {\em TESS} missions were
used to construct periodograms. By inspecting the light curves and
periodograms and knowing the approximate location of a star in the \mbox{H--R}
diagram from its effective temperature or spectral type, a variability
class can be assigned.  The results described here are from {\em TESS} 
sectors 1--68.

DSCT stars are defined to be main sequence F or A stars 
($6000 < T_{\rm eff} < 10\,000$\,K) which pulsate with high frequencies.  In 
this respect, ``high frequency'' is taken to mean any frequency higher than 
5\,d$^{-1}$. Lower frequencies may, or may not, be present.  If only
frequencies less than 5\,d$^{-1}$ are present, the star is classified as
GDOR.

Stars with frequencies less than 5\,d$^{-1}$ and in the temperature range 
$10\,000 < T_{\rm eff} < 18\,000$\,K are classified as SPB.  SPB variables also
include stars with $T_{\rm eff} > 18\,000$\,K, but in this temperature range
the maximum frequency is restricted to less than 3\,d$^{-1}$.  This is
necessary because BCEP stars may have frequencies as low as 3\,d$^{-1}$. 

Stars with $T_{\rm eff} > 18\,000$\,K and at least one frequency peak higher 
than 3\,d$^{-1}$ are defined as BCEP.  BCEP stars with low frequencies are 
classified as BCEP+SPB hybrids, but this may not be necessary as most BCEP 
stars are ``hybrids'' anyway.

Main sequence stars with $10\,000 < T_{\rm eff} < 18\,000$\,K and frequencies 
higher than 5\,d$^{-1}$ are defined as MAIA variables.  It is not
possible to distinguish between MAIA and DSCT stars without introducing a
temperature boundary.  The value of $T_{\rm eff} = 10\,000$\,K is convenient,
being the boundary between the A- and B-type stars.

\begin{figure*}
\begin{center}
\includegraphics{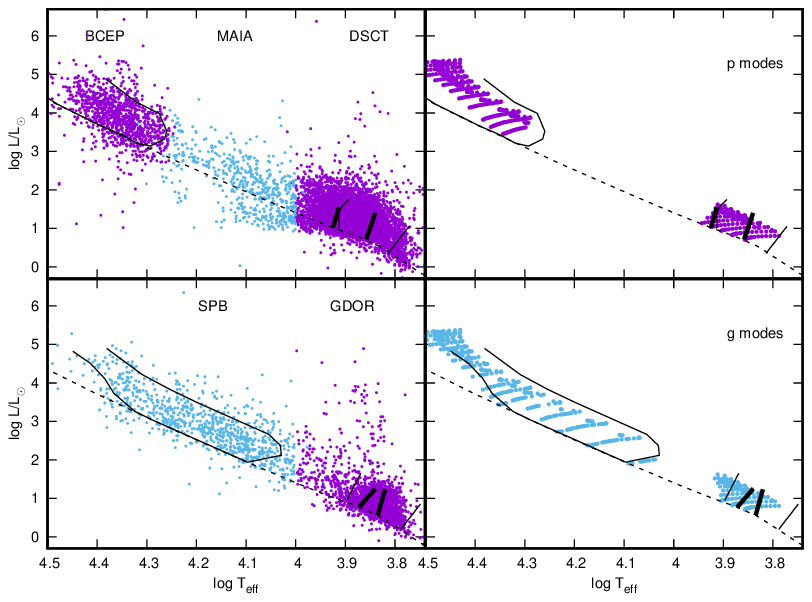}
\caption{The location of various classes of pulsating stars in the \mbox{H--R}
diagram (left panels) and location of Dziembowski models showing unstable p 
and g modes (right panels).  The dashed line is the zero-age main sequence.  
Also shown are the instability regions in BCEP and SPB stars for solar abundance 
models by \citet{Miglio2007b} and the hot and cool edges of the DSCT and GDOR 
stars from \citet{Dupret2005} (thick lines) and \citet{Xiong2016} (thin lines).} 
\label{hrdiag}
\end{center}
\end{figure*}

\begin{figure}
\begin{center}
\includegraphics{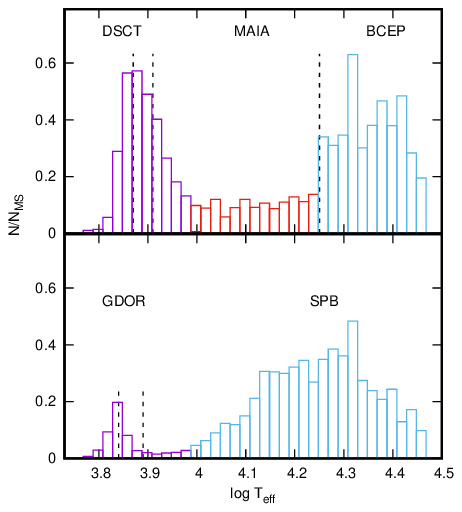}
\caption{The ratio $N/N_{\rm MS}$ as a function of effective temperature. $N$ is
the number of stars of the particular variability class and $N_{\rm MS}$
is the number of main sequence stars within the temperature bin size.  The
vertical dashed lines indicate the minimum and maximum temperature of the
hot edge of DSCT and GDOR stars predicted by various models.  The cool edge
of the BCEP instability region from \citet{Miglio2007} is shown.}
\label{vdensity}
\end{center}
\end{figure}

In addition to {\em TESS} photometry, linear non-adiabatic models were
calculated so that the predicted frequencies and location in the H--R
diagram could be compared with the observations.  Evolutionary stellar models 
were computed using the Warsaw - New Jersey evolution code 
\citep{Paczynski1970}, assuming an initial hydrogen fraction, $X_0 = 0.70$ and 
metal abundance, $Z = 0.020$ and using the chemical element mixture of 
\citet{Asplund2009} and OPAL opacities \citep{Rogers1992}.  Overshooting from 
the convective core was not included.   A mixing length parameter 
$\alpha_{\rm MLT} = 1.0$ was adopted for the convective scale height. 
All models are non-rotating.  The non-adiabatic code developed by 
\citet{Dziembowski1977a} was used to obtain the pulsation frequencies and
growth rates.  These models will be called the ``Dziembowski'' models.

\section{Location in the \mbox{H--R} diagram}

Comparison of the location and extent of the observed and predicted instability
strips is an important test of the models. The location in the H--R diagram
of BCEP, SPB, DSCT, and GDOR stars is shown in the left panels of
Fig.\,\ref{hrdiag}.  Also shown are the theoretical instability strips of
BCEP and SPB stars from \citet{Miglio2007} and the hot and cool edges of the 
DSCT and GDOR instability strips from \citet{Dupret2005} and \citet{Xiong2016}.
The panels on the right show the location in the H--R diagram of the
Dziembowski models with unstable modes as well as the theoretical
instability strips. 

The most obvious discrepancy is existence of the MAIA class of variables which
is not predicted by the models.  The large number of DSCT stars hotter than the 
theoretical hot edge of the instability strip also presents a serious problem.
The continuous band of low-frequency main sequence stars between the cool end 
of the SPB instability strip and the hot end of the GDOR instability strip 
should also be noted.

The number density of stars decreases rapidly with effective temperature. 
Whereas F0 stars are very common, B0 stars are very rare.  Fig.\,\ref{hrdiag}
therefore gives a somewhat biased view of the relative number of stars of
different variability classes. The fraction of stars of a given variability 
class relative to the number of main sequence stars at a particular temperature
is shown in Fig,\,\ref{vdensity}.  

Over 500 MAIA variables have been classified from {\em TESS} photometry.  The 
figure shows that MAIA variables comprise more than 10\,\% of mid- to 
late-B stars and are relatively more numerous than GDOR stars. The figure also 
shows that the large majority of main sequence stars do not seem to pulsate at 
all, or at least do not pulsate with amplitudes that can be detected by 
{\em TESS}.  This, too, is not understood.  

Fig.\,\ref{vdensity} shows how poorly the theoretical hot edge of DSCT
stars agrees with observations.  It also shows how the relative numbers of 
DSCT stars match smoothly with those of MAIA variables.  The MAIA stars seem
to be just an extension of the DSCT variables.  The figure also illustrates the
continuity between GDOR and SPB stars.

\begin{figure}
\begin{center}
\includegraphics{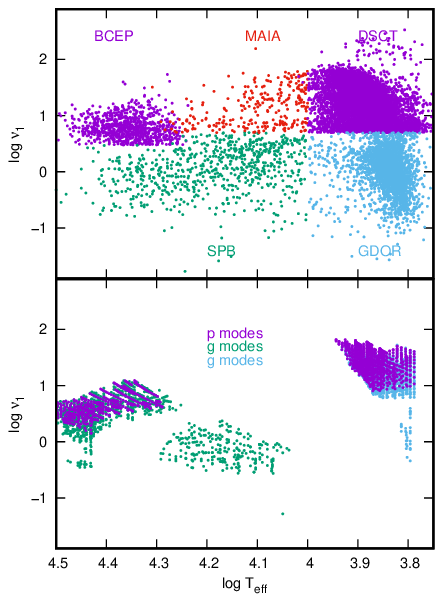}
\caption{The top panel shows the frequency of largest amplitude of various
types of {\em TESS} pulsating stars as a function of effective temperature.
The bottom panel shows the frequencies of unstable modes with $l \le 2$ in
non-rotating Dziembowski models.}
\label{nut}
\end{center}
\end{figure}

\begin{figure}
\begin{center}
\includegraphics{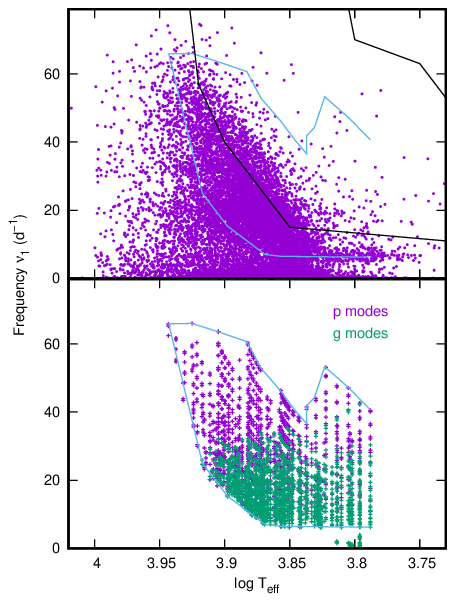}
\caption{The top panel shows the frequency of maximum amplitude as a function 
of effective temperature for DSCT stars.  The black outline is the envelope
of the highest amplitude growth rate of low-degree p modes shown in Fig.\,9 
of \citet{Xiong2016}. The blue outline is the envelope of unstable modes
from Dziembowski models.  The frequencies of unstable modes in the
Dziembowski models is shown in the bottom panel.}
\label{scud}
\end{center}
\end{figure}

\section{Pulsation frequencies}

In the previous section, it was shown that pulsation models fail 
to correctly predict the hot edge of the DSCT instability region, the MAIA
stars or the large number of low-frequency pulsators between the SPB and
GDOR instability strips.  In this section, the frequencies of unstable modes
from models are compared to the observed frequencies.

In the top panel of Fig.\,\ref{nut}, the frequency of largest amplitude,
$\nu_1$, is shown as a function of effective temperature.  This may be 
compared with the frequencies of unstable modes with $l \le 2$ in non-rotating 
Dziembowski models (bottom panel).  There are important discrepancies 
between observations and the models.  The difference between the observed and 
predicted frequencies in GDOR stars is understandable because the adoption of 
frozen-in convection in the models is not very satisfactory.  This should not 
affect models of the hotter stars where convection may perhaps be less 
important.  The models do not show the low frequencies which are present in
DSCT stars hotter than about 7500\,K.  The models also fail to reproduce the 
low frequencies in BCEP stars.  The distinctive frequency gap between the BCEP 
and the SPB stars in the models does not exist.

The top panel of Fig.\,\ref{nut} shows a surprisingly well-defined upper
frequency envelope in DSCT stars.  The frequency of maximum amplitude, $\nu_1$,
increases with temperature, reaching a maximum at about $\log T_{\rm eff} 
\approx 3.93$ before decreasing.  The decrease in $\nu_1$ continues in the
MAIA stars.   This is shown in more detail for the DSCT stars in the top
panel of Fig.\,\ref{scud}.  The bottom panel shows unstable p and g modes from 
the Dziembowski models.  The schematic envelope in this figure is reproduced in
the top panel (blue line).  Also shown in the top panel is the schematic 
envelope of p-mode frequencies with the highest growth rate from Fig.\,9 of 
\citet{Xiong2016} (black line).  

The frequencies of DSCT stars derived by \citet{Xiong2016} are in very poor 
agreement with observations. The frequencies from the Dziembowski models 
are in somewhat better agreement, but far from satisfactory.  For $\log 
T_{\rm eff} > 3.87$, low frequencies are stable in the Dziembowski models, 
whereas observations show that DSCT stars with low frequencies are the rule 
rather than the exception, as already noted above.

It does not matter whether the frequency of the largest amplitude, $\nu_1$, or 
second-largest amplitude, $\nu_2$, or the frequency with the $n$-th 
largest amplitude, $\nu_n$, is used.  The upper envelope of frequency as a 
function of effective temperature is always surprisingly sharp.  The
frequency of the envelope at constant temperature increases slowly with $n$.

If, instead, the maximum frequency, $\nu_{\rm max}$, is plotted as a function 
of effective temperature, the result is similar to Fig.\,\ref{scud}, except 
that the envelope is shifted towards higher frequencies by about 20\,\%.  
This is still considerably lower than the estimated critical frequency,
which is the frequency beyond which the pulsational acoustic waves are no 
longer reflected in the atmosphere.  

\section{Is the GDOR class necessary?}

The frequency threshold of $\nu_t = 5$\,d$^{-1}$, used to distinguish between 
DSCT and GDOR stars has no basis in theory.  It was adopted because 
frequencies lower than 5\,d$^{-1}$ were never seen in ground-based photometry 
of DSCT stars. This is not surprising because variations in atmospheric 
extinction and low pulsation amplitudes make the detection of low frequencies 
very difficult from the ground.

The detection of multiple low-frequency peaks in GDOR stars was unexpected
and not predicted from the pulsation models available at that time.  For
this reason, it was presumed to be a new class of pulsating variable
\citep{Balona1994b}.  Many more GDOR stars were subsequently discovered 
\citep{Henry2007}.  However, there was no similar attempt to detect low 
frequencies in the DSCT stars known at the time.

The first example of ``hybrid'' DSCT+GDOR pulsation was discovered 
by \citet{Handler2002b}.  Until the advent of space photometry, only 6 hybrids 
and 66 GDOR stars were known.  Surprisingly, the first {\em Kepler} release 
revealed that most DSCT stars were hybrids \citep{Grigahcene2010}.  
Indeed, among the {\em TESS} DSCT stars, about 75\,\% are hybrids.  

In retrospect, it seems that if sufficient photometric precision were 
available, ground-based observations would have detected low frequencies in 
most DSCT stars at an early stage.  In that case, the definition of DSCT
stars would not have excluded low frequencies.  What we now call GDOR stars 
would not have been a surprise and most likely regarded simply as DSCT stars 
with low frequencies.

\citet{Guzik2000} proposed the convective blocking mechanism to explain the
low frequencies in GDOR stars.  Later, \citet{Dupret2005} and \citet{Xiong2016} 
showed that the GDOR stars as well as low frequencies in the cooler DSCT
stars could be modeled using a time-dependent convection theory.  
\citet{Xiong2016} concluded that there is no essential difference 
between DSCT and GDOR stars. They can be considered as just two subgroups of 
one broader variability class.  

\begin{figure}
\begin{center}
\includegraphics{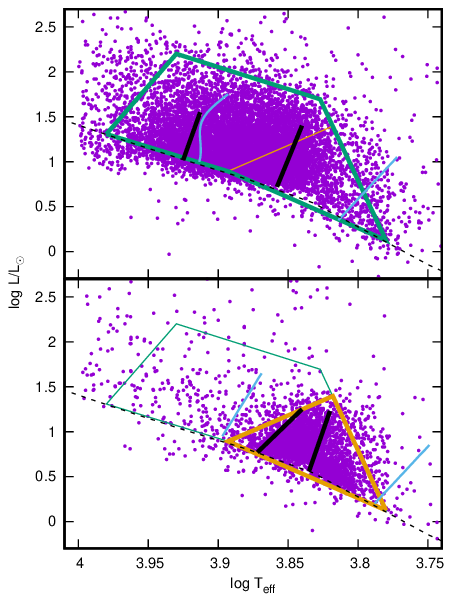}
\caption{The top panel shows the location of 12772 {\em TESS} DSCT stars in the
\mbox{H--R} diagram enclosed by a schematic trapezium (green). The hot and cool edges 
from \citet{Dupret2005} (thick black lines) and \citet{Xiong2016} (thin blue 
lines).  The bottom panel shows 4065 {\em TESS} GDOR stars enclosed by
schematic triangle (orange) and corresponding hot and cool edges. The 
diagonal dashed line is the zero-age main sequence.}
\label{scudor}
\end{center}
\end{figure}

\begin{figure*}
\begin{center}
\includegraphics{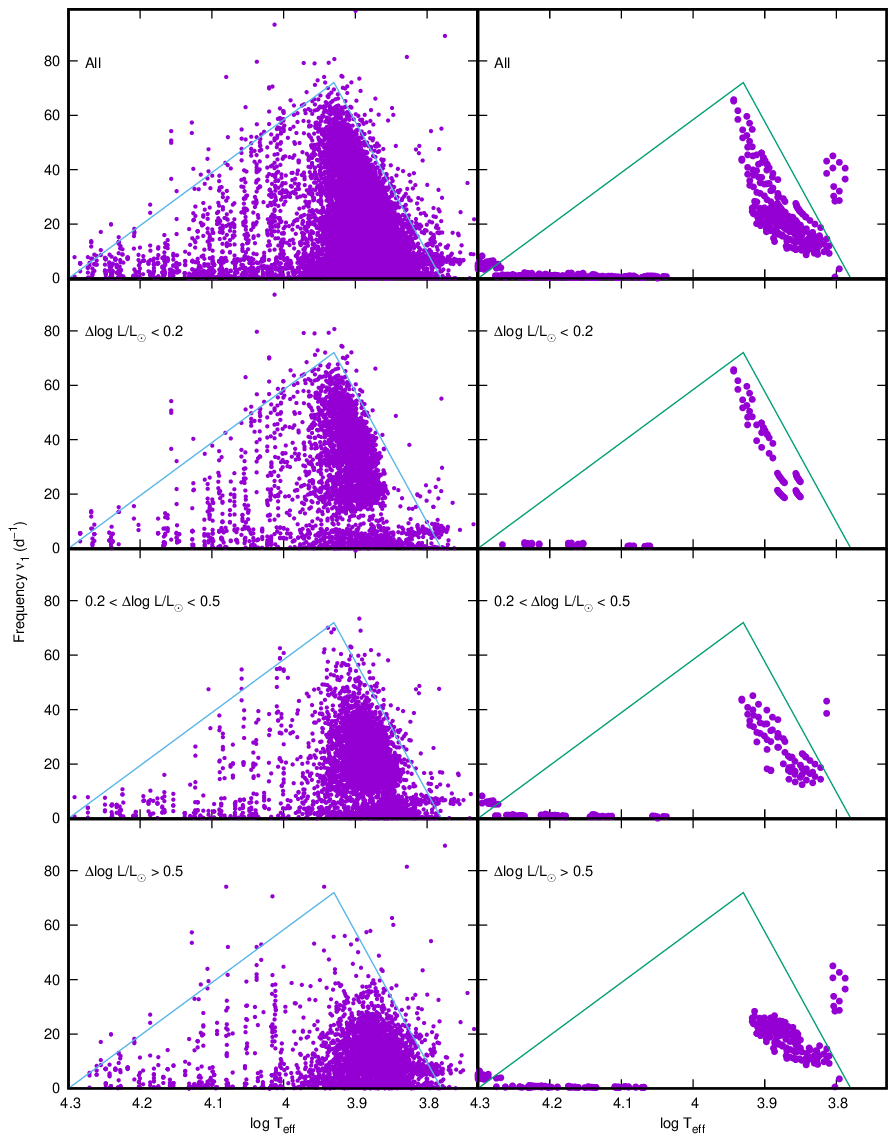}
\caption{The frequency of maximum amplitude as a function of effective 
temperature.  The top left panel shows all DSCT and MAIA stars enclosed
by a triangular envelope.  The other panels on the left show 
stars selected according to $\Delta L/L_\odot$, the luminosity above the 
zero-age main sequence.  The panels on the right are the corresponding
frequencies from the Dziembowski models.}
\label{fevol}
\end{center}
\end{figure*}

Since there seems no need to differentiate between DSCT and GDOR stars in the
models, it seems appropriate to re-evaluate the relevance of a separate GDOR 
class. Fig.\,\ref{scudor} shows the  DSCT and GDOR stars in the 
\mbox{H--R} diagram.  For convenience, the boundary of the DSCT and GDOR stars is 
approximated by a trapezoid (green) and triangle (orange) respectively.  Most 
GDOR stars are located inside the triangular region on the cool side of the 
DSCT stars, more or less within the region of low-frequency g-mode instability 
predicted by \citet{Xiong2016}.  The models by \citet{Dupret2005} predict a 
much smaller region.

There are 3995 DSCT stars and 3237 GDOR stars inside the triangular GDOR
region.  In other words, there are as many DSCT stars as GDOR stars.  The 
presence of GDOR and DSCT stars occupying the same region of the \mbox{H--R} 
diagram is impossible to duplicate in pulsation models.  Furthermore, there is 
a large number of GDOR stars outside the main triangular region, extending 
well beyond the hot edge of DSCT stars.  This, too, cannot be explained by
current models.  Furthermore, there are 35\,792 {\em TESS} main sequence stars 
within the triangular GDOR region that are not observed to pulsate at 
all.  The presence of stars with high frequencies, stars with low
frequencies, and ostensibly non-pulsating stars in the same region of the \mbox{H--R} 
diagram is not understood.  Models predict that all stars should pulsate.

Suppose that a threshold frequency, $\nu_t$, is used to discriminate between
GDOR and DSCT stars.  Let $n$ be the number of significant frequency peaks
below $\nu_t$. If there is any significance to $\nu_t$, then one might hope
to detect a difference in the distribution of $n$ as a function of $\nu_t$.
When this procedure is carried out with $\nu_t = 4, 6, 8$ and $10$\,d$^{-1}$, it
is found that the distributions of $n$ are very similar.  This suggests that
there is no reason to regard stars with low frequencies as different from
normal DSCT stars.  

Recently, it has been established that there is an extraordinary range in
frequency patterns among stars classified as DSCT \citep{Balona2024a}.  This
is at variance with the linear assumption which underlies all current 
pulsation models. The fact that DSCT and GDOR stars can be found in the same 
region of the H--R diagram is probably a result of non-linear mode excitation.

\section{Connection between DSCT and MAIA}

\citet{Balona2023a} has shown that there is no difference in mean rotation 
rates of MAIA stars and normal main sequence stars and that the amplitude 
distribution of MAIA variables is the same as in DSCT stars, but different
from BCEP stars. One may as well regard MAIA stars as an extension of DSCT
variables to hotter temperatures.  However, this conclusion awaits a theory
that explains pulsation in DSCT stars hotter than the currently predicted
limit as well as unstable high-frequency modes in late- to early-B stars.

The left panels of Fig.\,\ref{fevol} show $\nu_1$, the frequency of maximum 
amplitude, as a function of effective temperature for DSCT and MAIA stars.  
For MAIA stars, the plot shows $\nu_1$--$\nu_5$, the frequencies of the five 
highest amplitude peaks.  This is to balance the large difference in number
density between the two classes. 

The top left panel of Fig.\,\ref{fevol} shows that there is a roughly linear
increase in $\nu_1$ from the coolest DSCT stars, reaching a maximum of about 
72\,d$^{-1}$ at $\log T_{\rm eff} \approx 3.93$.  Thereafter the frequency 
decreases with temperature, reaching a minimum at the hot end of the MAIA
variables, as shown by the blue lines in the figure.  The systematic
decrease in $\nu_1$ initiated in the hot DSCT stars and continued by the
MAIA variables seem to be another link between DSCT and MAIA stars.

The top right panel of Fig.\,\ref{fevol} shows the frequencies 
$\nu_1$, $\nu_2$, of the first two unstable modes of highest growth rate and
visibility as a function of temperature from the Dziembowski models.
Not surprisingly, nearly all the modes in this plot are radial 
with radial order increasing with frequency from 1 to 8.  The increase in 
frequency with increasing temperature is also visible in the models.

Fig.\,\ref{fevol} also shows the $\nu_1$--$\log T_{\rm eff}$ diagram for
stars in different luminosity ranges, $\Delta L/L_\odot$, above the zero-age 
main sequence (ZAMS).  As  $\Delta L/L_\odot$ increases, the frequency of
maximum amplitude decreases from 72\,d$^{-1}$ near the ZAMS to about 
40\,d$^{-1}$ for the most evolved stars.  The corresponding effective 
temperature at maximum frequency decreases from 8500\,K for stars close to the 
ZAMS, to about 7500\,K for more evolved stars.  A similar behaviour seems to 
be present in the Dziembowski models (right panels in the figure).

The relationship between effective temperature and frequency in DSCT stars
has been investigated by many authors.  \citet{BarceloForteza2018,
BarceloForteza2020} found a linear relationship between an amplitude-weighted 
frequency and $T_{\rm eff}$ in DSCT stars.  \citet{Bowman2018} found a similar
relationship for the frequency of maximum amplitude.  However, as can be
seen from the above figures, there is no unique relationship between $\nu_1$
and $\log T_{\rm eff}$.

\begin{figure}
\begin{center}
\includegraphics{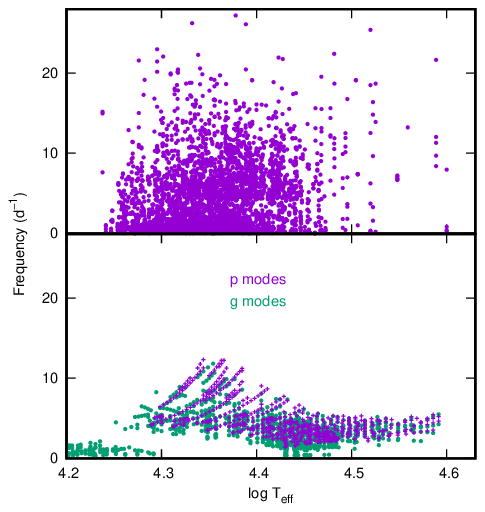}
\caption{All observed frequencies in BCEP, BCEP+SPB or pure SPB stars in the
BCEP temperature range are shown in the top panel.  The bottom panel shows 
unstable modes for $l\le 2$ from Diembowski models.}
\label{bcepfrq}
\end{center}
\end{figure}

\section{The SPB and BCEP stars}

Unless colours in the $U$ band are available, it is not possible to distinguish
between the effect of temperature and interstellar reddening in B stars.  
Since modern CCDs are not sensitive in the $U$ band, one has to
rely mostly on spectroscopic estimates of effective temperature or estimates
from Str\"{o}mgren or Geneva photometry.   The proportion of B stars with 
reliable effective temperatures is therefore considerably smaller than in A or F 
stars.  The spectral type and luminosity class often provide the only 
reliable estimate of $T_{\rm eff}$ for B stars.  Spectroscopy of B stars,
such as those by \citet{Burssens2020} and \citet{Shi2023} are very
important.

The defined lower bound in effective temperature for SPB stars, $T_{\rm eff} =
10\,000$\,K, is less than the cool edge derived from models of non-rotating
SPB stars by about 1000\,K.  It has been shown that SPB stars cooler than the 
theoretical cool edge do not rotate at a significantly higher rate than normal 
main sequence stars \citep{Balona2023a}.  Rapid rotation cannot therefore be 
used to explain the presence of SPB stars beyond the cool edge.  Stars
with frequencies typical of SPB or GDOR stars occur all along the main
sequence.  Models fail to reproduce this continuous sequence of
low-frequency pulsators.

In the catalog of \citet{Stankov2005}, there are 93 confirmed BCEP
variables.  An additional 103 BCEP stars are found in the photometric survey
by \citet{Pigulski2008b}. To these can be added the 86 new BCEP variables
discovered by \citet{Labadie-Bartz2020}, for a total of 282 known BCEP stars.
In \citet{Balona2022c}, 800 new BCEP stars have been detected, bringing the
total to 1082 BCEP stars identified from {\em TESS} photometry. Of these,
621 are classified as BCEP+SPB hybrids.

Hybrid BCEP+SPB pulsators were first discovered from ground-based photometry
\citep{Jerzykiewicz2005, Handler2006a,Chapellier2006} and many more from
space photometry \citep{Degroote2010b, Balona2011b}.  From {\em TESS} data,
57\,\% of BCEP stars are classified as BCEP+SPB hybrids.

The problem with BCEP+SPB hybrids is not merely a problem of classification.
According to models, BCEP stars should not pulsate with frequencies lower
than about 3\,d$^{-1}$.  This is because g modes have high amplitudes in 
the core and are heavily damped.  The structure of the eigenfunction does 
play a role and, for some modes, driving can exceed damping in the inner 
layers.  

To model the low frequencies requires an increase in opacity by 
about a factor of four \citep{Pamyatnykh2004}.  The discrepancy between the 
models and observations may also be reduced, to some extent, by a choice
of OP opacities rather than OPAL opacities \citep{Dziembowski2008,Miglio2007}. 
Enhanced iron opacity to address this problem is discussed by 
\citet{Moravveji2016}.

The problem can be seen in the bottom panel of Fig.\,\ref{nut} or the bottom
panel of Fig.\,\ref{bcepfrq}.  There is a distinct frequency gap between
non-rotating models of SPB stars and BCEP stars, whereas no such gap appears
in the observations.  There are no unstable frequencies less than 3\,d$^{-1}$ 
for BCEP stars cooler than $\log T_{\rm eff} \approx 4.40$. 

Rapid rotation tends to lower pulsation frequencies and may also play a part 
in explaining the BCEP+SPB hybrids. In that case, pure BCEP stars should 
rotate more slowly than BCEP+SPB hybrids.  From 123 pure BCEP stars 
$\langle v\sin i\rangle = 113 \pm 8$\,km~s$^{-1}$ and from 354 BCEP+SPB stars 
$\langle v\sin i\rangle = 158 \pm 6$\,km~s$^{-1}$.  The difference in rotation 
rate is not very large and more evidence is required.

The top panel of Fig.\,\ref{bcepfrq} shows the complete set of frequencies 
observed in BCEP, BCEP+SPB, and pure SPB stars in the BCEP temperature range
as a function of effective temperature.  The upper envelope is quite well 
defined, reaching a maximum frequency of about 18\,d$^{-1}$ at $T_{\rm eff} 
\approx 22\,000$\,K.   However, frequencies in the non-rotating models 
(bottom panel) do not exceed 13\,d$^{-1}$.  In spite of this discrepancy, 
BCEP stars show the best agreement with pulsation models.

\section{Some unusual BCEP stars}

TIC\,115642252 (NGC\,1960\,109).  The field of this B2I 10.7\,mag star is 
rather crowded and likely contaminated by TIC\,115642245 (9-th magnitude). 
Nevertheless, this will not explain three high-frequency peaks (60.996, 
57.189 and 53.875\,d$^{-1}$) present in TIC\,115642252 but absent in
TIC\,115642245.  \citet{Anders2022b} list a photometric effective temperature 
of 11\,174\,K, which conflicts with the spectral type and is possibly unreliable.

TIC\,119462263 (HD\,133518) is a He-strong star (B2IVp) \citep{Zboril2000} 
with high magnetic field strength \citep{Alecian2014}.  The pulsation 
frequencies are 57.391 and 54.169\,d$^{-1}$ with a possible third peak at 
60.727\,d$^{-1}$.  These frequencies are much higher than generally seen 
in BCEP stars.  Although the field is quite crowded, the star is very bright 
($V = 6.4$\,mag) and other stars within a 2\,arcmin field of view are fainter 
than 10-th magnitude.

TIC\,304425262 (HD\,103079) is a very bright ($V = 4.9$) B4IV star. Apart
from frequencies that qualify it as BCEP+SPB, there is also an interesting set 
of 8 peaks between 55 and 62\,d$^{-1}$.  The possibility that these are 
artifacts owing to the star's brightness cannot be ruled out.

\section{Conclusions}

In this paper, the observed regions of instability in the \mbox{H--R} diagram 
for various classes of pulsating stars and their frequencies are compared with 
predictions from pulsation models. It is concluded that, except possibly for 
BCEP stars, the models are in poor agreement with observations.

The origin of the high frequencies in MAIA variables has frequently been
dismissed as due to rotation, even though \citet{Balona2020a} have shown that
the projected rotational velocities of MAIA stars are no different from main
sequence stars in the same effective temperature range.  More recently,
\citet{Balona2023a} has shown that the $v \sin i$ distribution of 145 MAIA 
stars agrees with the $v \sin i$ distribution of 6538 main sequence stars
in the same temperature range.  In any case, rapid rotation will not
explain frequencies over 60\,d$^{-1}$ which occur in about 10\,\% of
MAIA stars. The amplitude distribution of MAIA variables is also the same as 
DSCT stars \citep{Balona2023a}, showing that MAIA stars are not DSCT variables 
in B-star binaries.  The driving mechanism for these hot DSCT stars is
unknown and should be considered a major unsolved problem.

Another major problem is that DSCT stars do not behave as predicted at a 
fundamental level.  It is taken for granted that stars with the same effective 
temperature, luminosity, and rotation rate should have the same frequencies.  
Moreover, slight changes in stellar parameters should lead to only slight 
changes in the pulsation frequency spectra.  Observations show very 
different behaviour. The frequency patterns in DSCT stars with similar 
parameters exhibit an extreme variety \citep{Balona2024a}.  This suggests that 
mode selection is a highly non-linear process mostly determined by local 
conditions.  

There are serious discrepancies between the observed frequencies and
frequencies of unstable modes in models of DSCT stars.  For example, the 
models of \citet{Xiong2016} predict frequencies that are far higher than 
observed. 

It is found that there is a well-defined envelope for the frequency of
maximum amplitude as a function of temperature.  The same relationship
is found if, instead of the frequency of maximum amplitude, $\nu_1$, the 
frequency of the $n$-th largest amplitude, $\nu_n$, is used.  The maximum 
frequency and the frequency range is smallest in cool DSCT stars, but increases 
smoothly with temperature. The frequency reaches  a maximum of about 72\,d$^{-1}$ 
at $T_{\rm eff} \approx 8500$\,K then decreases.  

The decrease in frequency and
frequency range with temperature continues in the MAIA variables, eventually 
reaching a minimum at about $T_{\rm eff} \approx 18\,000$\,K.  The highest 
frequencies of unstable modes in DSCT models do have an upper limit in 
reasonable agreement with observations.  This limit is related to the thermal 
timescale in the driving region.  However, models do not predict unstable high 
frequencies in stars hotter than about 8500\,K.  The decrease in maximum
frequency for stars hotter than this limit may provide a clue to the driving 
mechanism in MAIA variables.

The presence of equal numbers of DSCT and GDOR in the same region of the 
\mbox{H--R} diagram shows the pulsation mechanism must be the same in the two 
classes.  The arbitrary frequency used to separate DSCT and GDOR stars has no 
physical significance.  This seems to be just another example of the problem 
just described: very different frequency patterns can be produced by stars
with similar parameters.  Low frequencies are present in DSCT stars across the 
whole instability strip.  The reason why the GDOR class was created can be 
attributed to the difficulty of detecting low frequencies with small 
amplitudes from the ground.  Perhaps the GDOR class should be absorbed into 
the DSCT class. 

It has been suggested that SPB variables cooler than the cool edge of the
SPB instability strip can be explained by rapid rotation \citep{Szewczuk2012, 
Salmon2014}.  However, from 27 SPB stars cooler than 11\,000\,K, $\langle 
v\sin i\rangle = 184 \pm 20$\,km\,s$^{-1}$.  From 397 main sequence stars in 
the same effective temperature range, $\langle v\sin i\rangle = 124 \pm 
5$\,km\,s$^{-1}$.  The anomalous SPB stars do not seem to be rotating rapidly.

The failure of pulsation models to describe the observations requires a 
re-assessment of the basic assumptions underlying the models.  Perhaps
there is more than one driving mechanism active in upper main sequence
stars.  This might explain why DSCT and MAIA variables are seen over a very
wide range of effective temperatures.  It is also evident that a
re-assessment of the current variability classes is required.  However, this
must await future investigations on pulsational driving among hot main
sequence stars.

\section*{Acknowledgments}

I wish to thank the National Research Foundation of South Africa for 
financial support.

\section*{Data availability}

The data underlying this article can be obtained from

{\tt https://sites.google.com/view/tessvariables/home}.

and also available though the author.

\bibliographystyle{plainnat}
\bibliography{dsct}

\end{document}